\documentclass[aps,twocolumn,groupedaddress]{revtex4-1}

\usepackage{graphicx}% Include figure files
\usepackage{dcolumn}% Align table columns on decimal point
\usepackage{bm}% bold math
\usepackage{amsfonts}
\usepackage{amsmath}
\usepackage{amssymb}
\usepackage{xcolor}
\usepackage{enumerate}

\newcommand{\can}{Ca$_2$N}

\newcommand{\canii}{(CaN)$_2$}

\begin{document}

\title{Structural Transition in Oxidized \can\, Electrenes: CaO/CaN 2D 
heterostructures}

\author{Pedro H. Souza$^1$, Jos\'e E. Padilha$^2$, and Roberto H. Miwa$^1$}

\email{hiroki@ufu.br}

\affiliation{$^1$Instituto de F\'isica, Universidade Federal de Uberl\^andia,
        C.P. 593, 38400-902, Uberl\^andia, MG, Brazil}
        
\affiliation{$^2$Campus Avan\c cado Jandaia do Sul, Universidade Federal do 
Paran\'a, \\ 86900-000, Jandaia do Sul, PR, Brazil.}

\date{\today}
    
\begin{abstract}

Based on first-principles calculations we show that the oxidation of ultrathin 
films of \can\, electrides, electrenes, drives a 
hexagonal\,$\rightarrow$\,tetragonal structural transition. The ground state 
configuration of the oxidized monolayer (ML) and bilayer (BL) systems can be 
viewed as CaO/CaN and CaO/(CaN)$_2$/CaO two dimensional (2D) heterostructures. 
In both systems, we found nearly free electron (NFE) states  lying near the 
vacuum level, and the spatial projection reveals that they are localized above 
the oxidized CaO surface. Focusing on the magnetic properties, we find that the 
nitrogen atoms  of the oxidized \can\, becomes spin-polarized 
($\sim$1\,$\mu_{\rm B}$/N-atom); where (i)  the ferromagnetic  and the 
anti-ferromagnetic phases are nearly degenerated in the ML system, CaO/CaN, 
while (ii) there is an energetic preference for the  ferromagnetic phase  in 
CaO/(CaN)$_2$/CaO. We show that such a FM preference can be strengthened upon 
mechanical compression. Further electronic  structure calculations reveal that  
the FM CaO/\canii/CaO presents half-metallicity, where the metallic channels 
project (predominantly) on the N-$2p_{x,y}$ orbitals.  In addition to the total 
energy results, molecular dynamic and phonon spectra calculations have been done 
in order to verify its  thermal and structural stabilities. Those findings 
suggest that CaO/\canii/CaO is a quite interesting, and structurally stable, 2D 
FM heterostructure characterized half-metallic bands sandwiched by NFE states 
lying on the oxidized surfaces.

\end{abstract}

\maketitle

\section{Introduction}

Two dimensional (2D) materials have been the subject of numerous studies 
addressing technological applications as well as  fundamental issues. The 
seminal work on graphene\,\cite{novoselovScience2004} boosted the synthesis of 
other 2D systems. For instance, transition metal 
dichalcogenides,\cite{novoselovPNAS2005,chhowallaNatChem2013} obtained through 
exfoliation processes  of their layered parents, and  2D boron sheets 
(borophene) synthesized by bottom-up techniques on metal 
surfaces.\cite{mannixScience2015,fengNatChem2016} In parallel to the 
experimental realizations, computational simulations have played an important 
role performing a virtual engineering on 2D systems;  proposing new materials, 
and providing an atomic understanding of their physical/chemical 
properties.\cite{mounetNatNanotech2018,haastrup2DMat2018,schlederJPC2019} The 
combination of those efforts has contributed  to the current  revolution in the  
material science based on 2D structures.\cite{bhimanapatiACSNano2015} 

Further control on the physical properties of those 2D materials can be done by 
tuning structural  parameters like  (i) the number of stacked layers or  (ii) 
the equilibrium lattice structure; {\it e.g.} in (i) we have the 
linear/parabolic  energy dispersion in graphene 
monolayer/bilayer,\cite{ohtaScience2006controlling,castroPRL2007biased, 
zhangNat2009,netoRevModPhys2009} and in (ii) the semiconducting/metallic  
electronic structure of 2H/1T 
MoS$_2$.\cite{linNatNanotech2014,voiryChemSocRev2015,gaoJPhysChemC2015, 
guoNanoLett2015} More elaborate strategies have been proposed/applied in order 
to tailor the electronic properties of 2D systems, as the design  of  van der 
Waals (vdW) heterostructures by stacking combinations of 2D 
materials\cite{geimNatMat2007,novoselovScience2016,jinNatnanotech2018}, 
functionalization by foreign 
elements,\cite{boukhvalovJPC2009chemical,chenChemRev2012,yanChemSocRev2012} and 
synthesis of Janus structures.\,\cite{luNatNanotech2017,duongACSNano2017, 
dongACSNano2017,liJPhysChemLett2017}

%%%%%%% NFES %%%%%%%%%%%%%%
It is worth noting that low dimensional systems, like bundles of carbon 
nanotubes and fullerenes host nearly free electron (NFE) 
states.\cite{okadaPRB2000,zhaoACSNano2009,huNanoLett2010} In fact, NFE states 
have been firstly  detected  on metal surfaces through inverse photoemission  
spectroscopy experiments,\cite{johnsonPRB1983,reihlPRB1984,hulbertPRB1985} and 
confirmed  by first-principles 
calculations.\cite{kiejnaPRB1991,eguiluzPRL1992,silkinPRB1999,hillPRL1999, 
rhm_gpsSUSC2001} Such  NFE states are also present in 2D systems. For instance, 
in graphene  those states were measured by near edge x-ray spectroscopy, 
\cite{pacilePRL2008} and further supported by theoretical first-principles DFT 
calculations.\cite{silkinPRB2009} 

Two dimensional dicalcium nitride is a layered system which hosts NFE states.  
Each \can\, layer unit is composed by a nitrogen atomic sheet sandwiched by  
calcium layers connected by Ca--N--Ca chemical bonds, whereas the  \can\, layer 
units are stacked along the (001) direction bounded   by   van der Waals (vdW) 
interactions.  Based on the oxidation states of Ca$^{2+}$ and N$^{3-}$ atoms, we 
can infer that each \can\, unit is positively charged, [\can]$^{+}$, and thus in 
order to provide electrostatic stability to the  system,  negative charges 
(anionic electron)  take place between the  \can\, 
layers.\cite{leeNature2013,walshJMatChemC2013} The NFE states are partially 
occupied by those anionic electrons, giving rise to metallic bands for 
wave-vectors on the [001] plane.\cite{OhJACS2016} Meanwhile, in few layer 
systems, it has been predicted the presence of the anionic electrons not only 
embedded between the \can\, layers, but also on the surface region of the slab, 
giving rise to NFE surface states.\cite{inoshitaPRB2017} 

Few years ago, 2D nanosheet systems of \can\, electrides (electrenes) have been 
successfully synthesized through exfoliation processes.\cite{druffelJACS2016} In 
fact, the metallic character, ruled by ionic electrons, was already predicted in 
the  \can\,   monolayer systems.\cite{zhaoJACS2014}  Since the presence of 
metallic surface states makes the  \can\, electrenes chemically very 
reactive,\cite{druffelJMatChemC2017} it has been proposed  some strategies  
aiming to circumvent such a problem. For instance, the encapsulation of the 
\can\, nanosheet using 2D shields like graphene, single layer of BN, or 
graphane.\cite{zhaoJACS2014} On the other hand,  NFE states make the \can\, 
surface a quite interesting platform  to perform atomic  design on 2D 
materials,\cite{celottaRSI2014} for instance through  functionalization by 
foreign atoms or 
molecules.\cite{weeksPRX2011,gomesNature2012,acostaPRB2014,DrostNatPhys2017, 
SlotPRX2019,CrastoNanoLett2020}  Indeed, OH-functionalized  \can\, and Y$_2$C 
monolayers have been proposed as new 2D materials for application in ion 
batteries;\cite{wangASUS2019} and very recently, it has been proposed the 
control of the electronic and magnetic properties in \can\, monolayers through  
hidrogenation and oxidation processes.\cite{qiuJPhysChemC2019,wuJMMM2020}

Finally, it is well known that,  in order to minimize the surface free energy, 
the surface atoms may rearrange upon   incorporation of foreign elements. 
Such  surface reconstructions may result in  noteworthy electronic properties. 
For instance, the formation of NFE states and the nearly 1D metallic 
transport channels on the (single layer) indium covered  silicon surface, 
In/Si(111)-$(\sqrt{7}\times\sqrt{3})$.\cite{rotenbergPRL2003,yamazakiPRL2011} 

Once it is reasonable to infer that such a surface atomic rearrangements may 
take place in functionalized 2D systems; here, based on the density functional 
theory (DFT),   we performed a set of first-principles  calculations of the 
fully oxidized  monolayer (ML) and bilayer (BL) \can\, electrenes. Our total 
energy results reveal a hexagonal\,$\rightarrow$\,tetragonal structural  
transition in the oxidized  \can-ML and -BL, resulting in  2D tetragonal 
heterostructures composed by  stacked  layers of CaO and CaN, namely CaO/CaN and 
CaO/\canii/CaO, respectively. In the sequence, the perform a detailed study of 
the electronic and magnetic properties of those systems. We found that both 
oxidized electrenes are metallic. The  ferromagnetic phase  is strengthened in 
the  bilayer system; CaO/\canii/CaO is characterized by half-metallic bands 
localized within the \canii\, layers, sandwiched by NFE states lying on the 
oxidized CaO surface. Molecular dynamics and phonon spectra calculation were 
done in order to confirm the thermal and structural stability of CaO/\canii/CaO. 
Those findings suggest that CaO/\canii/CaO is a potential structure for 
application in 2D nanodevices.

\section{Computational details}

The calculations were performed by using  the density functional theory 
(DFT),\cite{kohn} as implemented in the computational codes Quantum-Espresso 
(QE)\,\cite{espresso} and Vienna Ab initio Simulation Package 
(VASP).\cite{vasp1,vasp2} We have considered the generalized gradient 
approximation of Perdew-Burke-Ernzerhof (GGA-PBE)\,\cite{PBE} for the 
exchange-correlation functional.  The Kohn-Sham\,\cite{KS} orbitals, and the 
self-consistent total charge density  were expanded in plane wave basis sets 
with energy cutoffs of 70 and 353\,Ry, respectively. The  Brillouin zone 
sampling was performed by using a  8$\times$8$\times$1 k-point mesh.\cite{mp} 
The \can-ML and -BL systems were described within the supercell approach with 
surface periodicities of (1$\times$1), ($\sqrt 2$$\times$$\sqrt 2$), and
(3$\times$3), and a vacuum region of between 22 and 28\,\AA. The atomic 
positions were relaxed until the residual forces were converged to within 
5\,meV/\AA, and the structural relaxation (variable-cell) was performed within 
a pressure convergence of 0.05\,Kbar. The long-range van der Waals (vdW) 
interactions were described using the semiempirical approaches vdW-D2 and 
-D3.\cite{grimme2006semiempirical,grimmeJChemPhys2010} We have also  checked 
the validity of our results using the self-consistent vdW-DF 
approach.\cite{dionPRL2004,perezPRL2009,klimevsPRB2011}

The structural stability was verified through the calculation of elastic 
constants and the phonon dispersion using PHONOPY code;\cite{togoScrMat2015}
and the thermal stability was verified by {\it ab initio} molecular 
dynamics simulations (AIMD) at 300\,K, with a time step of 1\,fs using Nos\'e 
heat bath scheme.\cite{noseJChemPhys1984}

  \begin{figure}
    \includegraphics[width=\columnwidth]{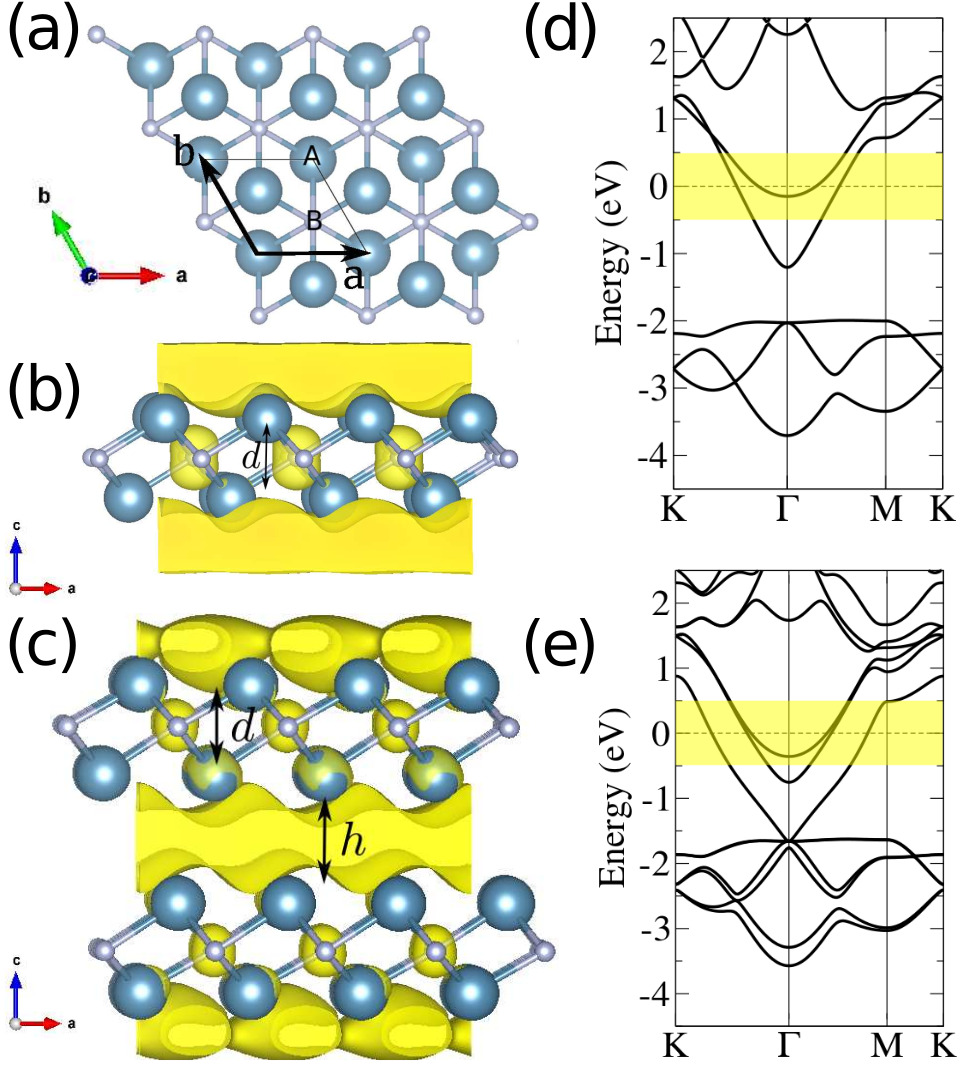}
    \caption{\label{models}Structural model of \can-1ML,  top-view (a), 
side-view and the partial  electron charge density  within an energy interval of 
$E_F$$\pm$0.5\,eV of \can-ML (b) and can-BL (c). Electronic band structure of 
\can-ML (d) and -BL (e). Isosurfaces of 
0.002 (b) and 0.005\,$e$/\AA$^3$.}
  \end{figure}

\section{Results and Discussions}

{\bf Pristine ML and BL \can.} In Fig.\,\ref{models} we present the structural 
models and the electronic structure of pristine \can-ML and -BL. In  \can\,  the 
N atoms are intercalated by Ca atoms, forming a triangular lattice, 
Fig.\,\ref{models}(a). At the equilibrium geometry, the \can-ML presents a 
lattice constant ($a$) of 3.56\,\AA, and  Ca--Ca vertical distance ($d$) of 
2.51\,\AA\, [Fig.\,\ref{models}(b)]. The  \can-BL presents practically the 
same values of $a$ and $d$, and interlayer distance $h$\,=\,3.62\,\AA, 
Fig.\,\ref{models}(c). The energetic stability of the bilayer system is dictated 
by vdW interactions, where we found an interlayer binding energy of 
79\,meV/\AA$^2$ (1.26\,J/m$^2$). The electronic band structures of 
\can-ML and -BL, Figs.\,\ref{models}(d) and (e), reveal the formation of 
parabolic metallic bands for wave vectors parallel to the \can\, surface layer. 
The real space projection of those metallic parabolic bands within an energy 
interval of $\pm$0.5\,eV with respect to the Fermi level, $E_F\pm0.5$\,eV, 
reveals the formation  nearly free electron (NFE) states spreading out on the 
surface of \can-ML and -BL, and sandwiched between \can\, layers in the BL 
system, Figs.\,\ref{models}(b)  and (c).

  \begin{figure}
    \includegraphics[width=\columnwidth]{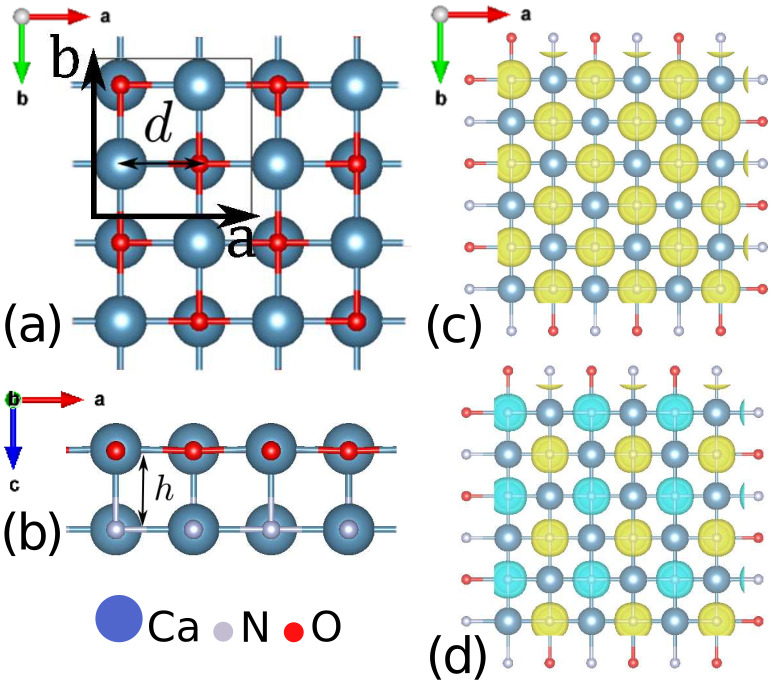}
    \caption{\label{models2a} Structural model of the fully oxidized \can-ML, 
forming tetragonal CaO and CaN stacked layers, CaO/CaN, top-view (a), and 
side-view (b). Spin-densities of the (c) FM and (d) AFM  CaO/CaN. The 
spin-densities are localized on the nitrogen atoms. Isosurfaces
of 0.005\,$e$/\AA$^3$.}
  \end{figure}
  
{\bf Oxygen adsorption.} Once we have characterized  the electronic and 
structural properties of pristine \can-ML and -BL, where we found  good 
agreement with the current literature,\cite{fangChemMat2000, leeNature2013, 
walshJMatChemC2013, zhaoJACS2014, OhJACS2016, druffelJACS2016, 
liuJPhysChemC2020} in the sequence we examine their functionalization by oxygen 
adatoms.

The energetic stability of the oxidized systems (O/\can) was inferred through 
the 
calculation of the oxygen adsorption energy ($E^a$), 
$$
E^a =  E[{\rm Ca_2N}] + E[{\rm O_2}]/2 - E[{\rm O/Ca_2N}],
$$ 
where $E[{\rm O/Ca_2N}]$   and $E[{\rm Ca_2N}]$  are the total energies of 
oxidized  and pristine \can\, (ML or BL), and $E[{\rm O_2}]$ is the total energy 
of an isolated O$_2$ molecule. Positive values of $E^a$ indicate exothermic 
processes. In addition, in  order to verify the emergence of 
magnetic phases, we compare the total energies of non-magnetic ($E^{\rm NM}$) 
and magnetic ($E^{\rm Mag}$) O/\can,   $\Delta E^{\rm mag} = E^{\rm NM} - E^{\rm 
Mag}$.

Firstly we calculate the adsorption energies of a single O adatom on the 
\can-ML. We have considered the sites A and B [Fig.\,\ref{models}(a)] in a 
(3$\times$3) surface, and we  found an energetic preference of 0.30\,eV for  the 
oxygen  adatom lying on the hollow site (A), which is aligned with the Ca atom 
at  the opposite side of the \can-ML.  We obtained $E^a$ of 4.94\,eV/O-atom, 
thus confirming  the  dissociative adsorption of O$_2$ molecules on the \can\, 
surface.\cite{zhaoJACS2014,liuMaterResExpress2018}

For a full coverage of oxygen adatoms, we found a structural transition of the  
oxidized \can-ML, O/\can-ML. We have considered the hexagonal lattice of 
pristine \can-ML as the initial  configuration, and upon structural and atomic 
relaxations, we found (i) a metastable phase where the lattice vectors {\bf a} 
and {\bf b} increase by 12.5\,\% with respect to the pristine \can-ML, and the 
planar angle between {\bf a} and {\bf b} ($\gamma$) increases from 120$^\circ$  
to 125$^\circ$, followed by (ii) a ground state tetragonal lattice with $|{\bf 
a}|$=$|{\bf b}|$=4.72\,\AA\, [Fig.\,\ref{models2a}(a)], and more stable than 
(i) by  0.97\,eV/(1$\times$1) with $E^a$\,=\,4.04\,eV/O-atom, pointing out a 
large thermodynamic preference for (ii). As shown in Fig.\,\ref{models2a}(b), 
O/\can-ML  is characterized by parallel layers of CaO and CaN, each one  forming 
planar square-lattices with Ca--O and Ca--N bond lengths of 2.36\,\AA\, [$d$ in 
Fig.\,\ref{models2a}(a)] connected by Ca--O vertical bonds [$h$=2.45\,\AA\, in 
Fig.\,\ref{models2a}(b)]. Such a tetragonal phase of O/\can-ML can be described 
as a layered  combination of trigonal structures of  CaO and  CaN  
crystals,\cite{MatProj}  forming a 2D CaO/CaN heterostructure. We can infer 
that the structural preference for the tetragonal phase is attributed to the 
larger formation energy of  CaO\,\cite{osti_1201098} compared with that of 
\can,\cite{osti_1201252} and also the proper stoichiometry upon the oxidation 
of one side of the  \can-ML, O/\can-ML. On the other hand, when both sides of 
\can-ML are  oxidized, O/\can-ML/O,  the hexagonal 
lattice and the atomic structure of \can-ML have been preserved.

  \begin{figure}
    \includegraphics[width=\columnwidth]{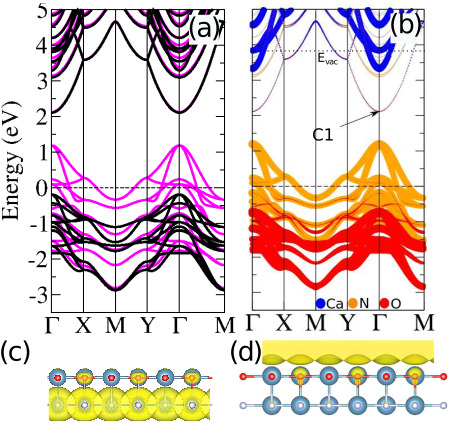}
    \caption{\label{models2b} (a) Spin-polarized electronic band structure of 
the FM-CaO/CaN, (b)  projection of the energy bands onto atomic orbitals; 
spacial projection of the metallic bands near the Fermi level 
($E_{\rm F}\pm 0.2\,\textrm{eV}$) (c), and the parabolic band ($c1$) near the 
$\Gamma$-point (d). Isosurfaces of 0.001\,$e$/\AA$^3$.}
  \end{figure}
  
Focusing on the magnetic properties, there is an energy gain of $\Delta E^{\rm 
mag}$=68\,meV/(1$\times$1)  upon the inclusion of spin-polarization. We found a 
net  magnetization  ($m$) of 1\,$\mu_{\rm B}$ localized on the N atoms, a shown  
in Figs.\,\ref{models2a}(c) and (d)  for the FM and AFM phases, respectively. 
The total energy comparison between   the  FM ($E^{\rm FM}$) and AFM ($E^{\rm 
AFM}$) phases indicates that the former is slightly more stable, with  $E^{\rm 
AFM}-E^{\rm FM}=7.6$\,meV/N-atom. Electronic band structure calculations 
indicate  the formation of metallic bands for both spin configurations (FM and 
AFM). The energy bands of the FM CaO/CaN [Fig.\,\ref{models2b}(a)], and the  
projection of the electronic states  near the Fermi level 
[$-0.15\,<E-E_F\,<\,0.5\,\textrm{eV}$ in Fig.\,\ref{models2b}(c)]  reveal a 
half-metallic system ruled by the N-$2p_{x,y}$ orbitals. Such a half-metallic 
behavior is suppressed in the AFM CaO/CaN. Further projection of the energy 
bands  onto the atomic orbitals of Ca, N, and O [Fig.\,\ref{models2b}(b)]  
indicates that there are no atomic orbital contributions to the lowest 
unoccupied parabolic band $c1$. Indeed, its spatial distribution near the 
$\Gamma$ point, spreading out on the CaO surface  [Fig.\,\ref{models2b}(d)], 
allow us to   identify   $c1$ as a NFE band localized at $\sim$1.6\,eV below the 
vacuum level ($\rm E_{vac}$).

  \begin{figure}
    \includegraphics[width=\columnwidth]{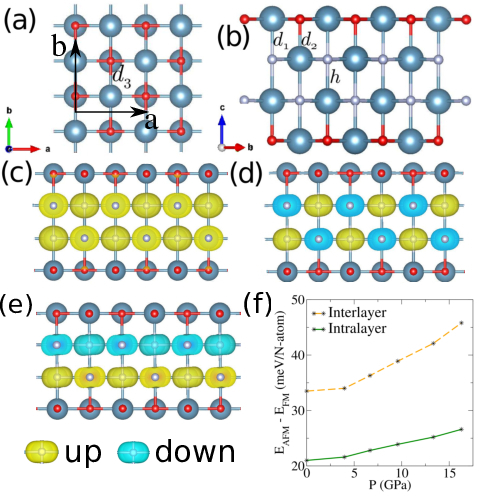}
    \caption{\label{models4a}Structural model of CaO/\canii/CaO, (a) top-view, 
and (b) side-view. The bond lengths $d_{i}$ ($i$=1, 2, and 3) are 2.47, 2.40, 
and 2.39\,\AA, respectively, and $h$\,=\,2.53\,\AA. Spin-densities of the  
intralayer/interlayer (c) FM/FM,  (d) AFM/AFM, and (e) FM/AFM  phases. (f) 
Total energy difference between the FM and AFM phases, of  the interlayer and 
intralayer couplings, as a function of the external (normal) compresive 
strain.  Isosurfaces = 0.005$\,e$/\AA$^3$.}  
  \end{figure}
  
We next have considered the oxidation of both sides of \can-BL, O/\can-BL/O.  
Similarly to what we have done in O/\can-ML,  we  took the hexagonal structure 
of pristine \can-BL as the starting configuration. By performing full structural 
and atomic relaxations, we found a tetragonal lattice, with  $|{\bf a}|$=$|{\bf 
b}|$=4.77\,\AA\, [Fig.\,\ref{models4a}(a)], more stable than the 
hexagonal lattice by 1.97\,eV/(1$\times$1). 
The oxidation process is exothermic by  $E^a$\,=\,4.14\,eV/O-atom, indicating 
that the  dissociative adsorption picture has been kept  for O$_2$ molecule  on 
the \can-BL surface. O/\can-BL/O can be  viewed as a combination of two CaO/CaN  
structures, however instead of vdW interaction, the formation of C--N chemical 
bonds, indicated as $h$ in Fig.\,\ref{models4a}(b), increases the interlayer 
binding energy from 79\,meV/\AA$^2$ (pristine \can-BL) to  93\,meV/\AA$^2$. The 
equilibrium geometry of O/\can-BL/O is characterized by an inner region that 
mimics the atomic structure of the trigonal CaN,   sandwiched by nearly planar 
layers of CaO, CaO/\canii/CaO.

\begin{figure}
    \includegraphics[width=\columnwidth]{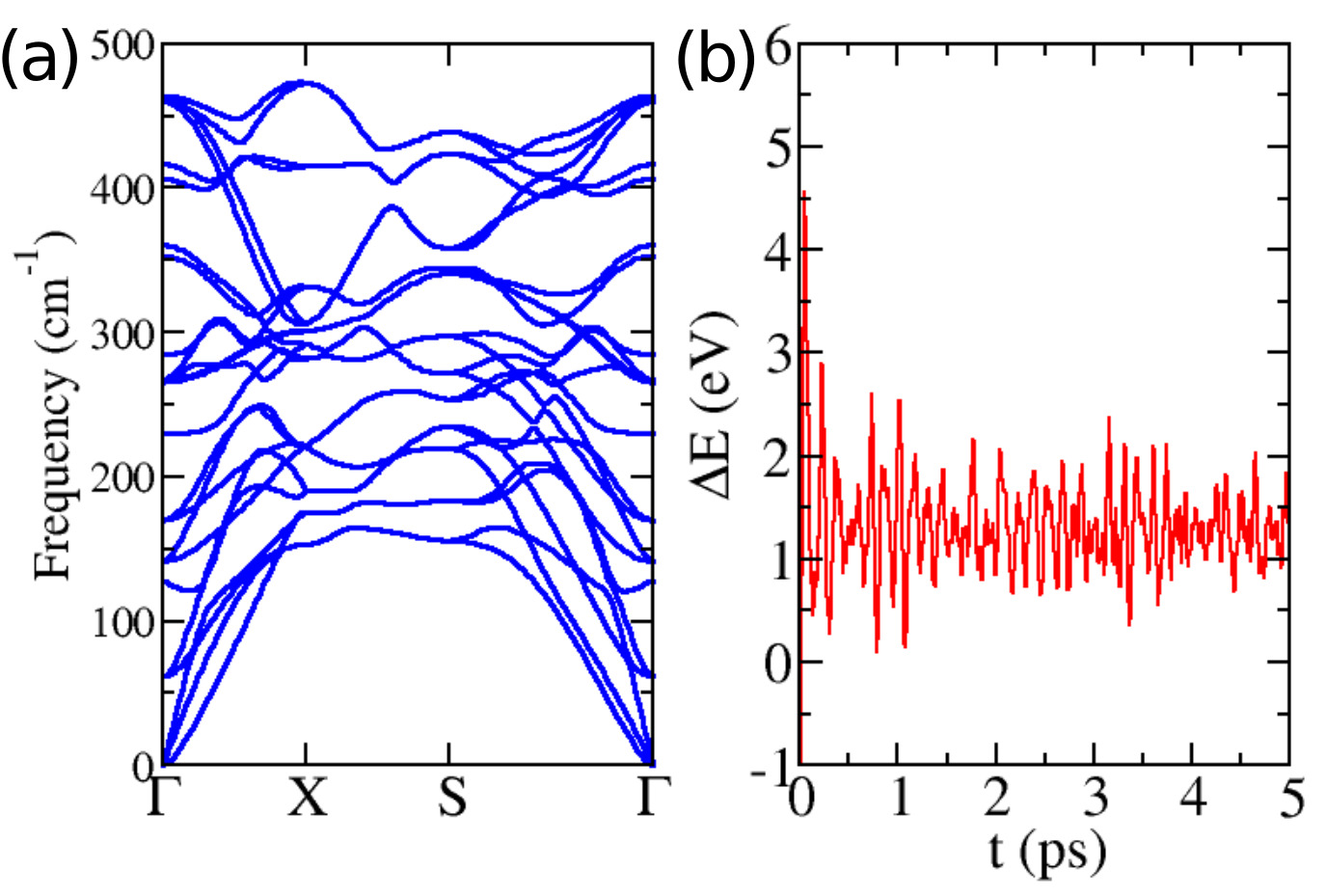}
    \caption{\label{phonon-md} (a) Phonon dispersion for O/\can-BL/O. (b) Energy 
Fluctuation at 300K. The insets, show snapshots of the crystal structures for a 
time of 5ps.}
  \end{figure}

In addition to the energetic stability of CaO/\canii/CaO, we have also examined 
its (i) structural, and (ii) thermal stabilities. The former [(i)], was 
performed by  the calculation of the phonon spectra, presented in 
Fig.\,\ref{phonon-md}(a).  We have considered the finite-displacement approach, 
with supercell sizes of ($3\times3$) primitive unit cells. As we can see, there 
are no  imaginary frequencies phonon frequencies. Those results allow us to 
infer that CaO/\canii/CaO is  structurally  stable. In (ii), we have performed 
{\it ab initio} molecular dynamics (AIMD) simulations at 300K, with a time step 
of 1 fs using the Nos\'e heating bath scheme.\cite{noseJChemPhys1984} We have 
constructed a ($3\times 3$) supercell containing 9 formula units to minimize the 
constraint induced by the slab periodicity. The results of the variations of the 
total potential energy with respect to the simulation time, and snapshots of the 
last configurations are presented in Fig.\,\ref{phonon-md}(b). As we can 
observe, the atomic configuration  remains nearly unaltered up to 5\,ps, and no 
phase transition is observed in the total potential energy. These results 
demonstrate that the 2D CaO/\canii/CaO material, once synthesized, is stable and 
preserving its structural integrity at room temperature.

The key aspects related to the magnetic properties of CaO/\canii/CaO are 
summarized in Figs.\,\ref{models4a}(c)--(f). There is an energy gain upon the 
inclusion of spin-polarization of $\Delta E^{\rm mag}$=70\,meV/N-atom, 
practically the same values obtained in CaO/CaN. Similarly, the  net 
magnetization is mostly localized on the N atoms, with $\sim$1$\mu_{\rm 
B}$/N-atom. Further comparison between the FM and AFM couplings revealed that, 
(i) there is an energetic preference of 21\,meV/N-atom for the FM configuration 
compared with the AFM configuration between the nitrogen atoms lying on the same 
CaN layer, intralayer coupling shown in Figs.\,\ref{models4a}(c) and (d). (ii)  
Between the stacked CaN layers, interlayer coupling,  we found the FM phase more 
stable than the AFM phase [Figs.\,\ref{models4a}(c) and (d)] by 32\,meV/N-atom. 
Here, the emergence of such a magnetic phase in  CaO/\canii/CaO can  can be 
considered as a remnant property of the (trigonal) CaN  crystal. Using the mean 
field theory, we estimate a Curie temperature ($T_{\rm C}$) in (i)  of 
162\,K.\cite{kudrnovskyPRB2004} 

It is worth noting that although nearly the same energy gain due to the 
spin-polarization ($\Delta E^{\rm mag}$), the the energetic preference for the  
intralayer FM configuration [(i)] has been strengthened in  CaO/\canii/CaO; {\it 
i.e.} $E^{\rm AFM} - E^{\rm FM}=7.6\rightarrow{21}$\,meV/N-atom. Such a 
preference can be attributed to the FM interlayer coupling [(ii)]. In this case, 
we can infer that the energetic  preference for the FM (interlayer and 
intralayer) phase  can be tuned by a compressive strain  normal to the 
CaO/\canii/CaO surface, and thus, increasing the Curie temperature. Indeed, as 
shown in Fig.\,\ref{models4a}(f), the total energy difference $E^{\rm AFM} - 
E^{\rm FM}$ of the intralayer and interlayer couplings increases upon external 
strain. For compression of 3\,\% normal to the stacking direction, which 
corresponds to an external pressure of 17\,GPa, the strength of the interlayer 
FM  coupling increases from 32 to 46\,meV/N-atom, resulting in an intralayer 
coupling of 27\,meV/N-atom, that is, $E^{\rm AFM} - E^{\rm 
FM}=21\rightarrow{27}$\,meV/N-atom, and $T_{\rm C}=162\rightarrow{209}$\,K.

% 
% due to the 
% FM interlayer coupling [(ii)],  the the energetic preference for the  intralayer 
% FM configuration has been strengthened in  CaO/\canii/CaO; {\it i.e.} $E^{\rm 
% AFM} - E^{\rm FM}=7.6\rightarrow{21}$\,meV/N-atom in (i).  Since the strength of 
% the FM phase in (i) is dictated by the interlayer coupling, (ii), we can infer 
% that such an energetic  preference of the FM phase  can be tuned by a 
% compressive strain  normal to the 2D CaO/\canii/CaO, and thus, increasing the 
% Curie temperature. Indeed, this is what we found by compressing the 
% CaO/\canii/CaO by 3\,\% normal to the stacking direction. The strength of the FM 
% phase ruled by the interlayer coupling increases from 32 to 60\,meV/N-atom.

  \begin{figure}
    \includegraphics[width=7.5cm]{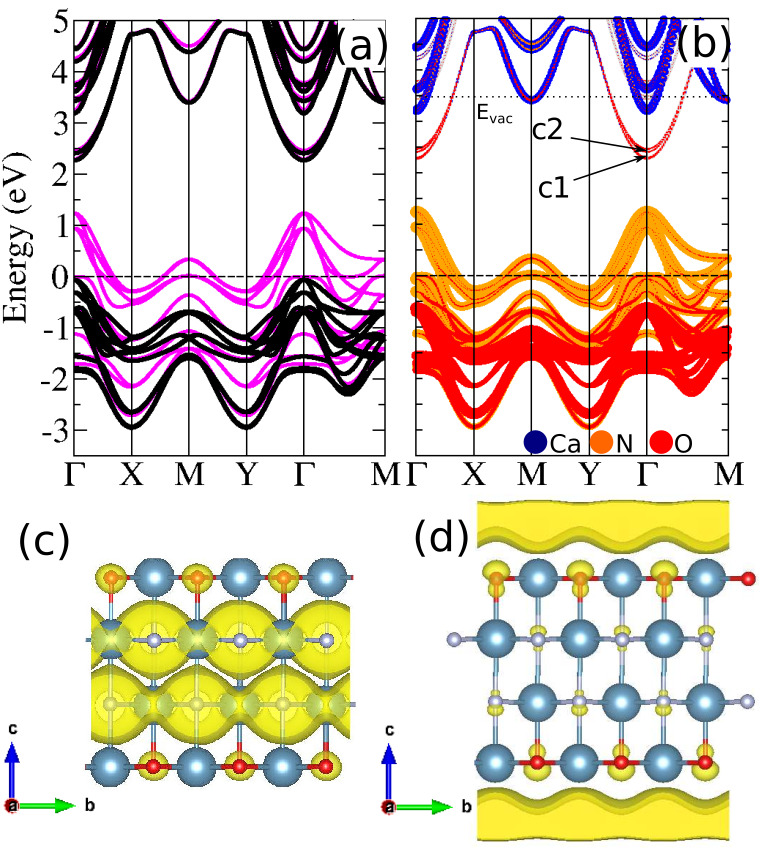}
    \caption{\label{models4b} Spin-polarized electronic band structure of the FM 
CaO/\canii/CaO (a), and spatial projection of the energy bands near the Fermi 
level ($E_{\rm F}\pm 0.2\,\textrm{eV}$ (c). (b) Projection of the energy bands 
on the Ca, N, and O atomic orbitals and spatial (d) the projection of the 
parabolic band $c1$. Isosurfaces of 
0.001$e$/\AA$^3$.}
  \end{figure}

The electronic band structure of the FM CaO/\canii/CaO characterizes  a 
half-metallic system,  where the spin-down metallic channels [purple lines in 
Fig.\,\ref{models4b}(a)] are (mostly) ruled by the N-$2p_{x,y}$ orbitals, 
Fig.\,\ref{models4b}(b). The charge density overlap of the metallic bands 
through the N atoms is depicted in Figure\,\ref{models4b}(c). Further energy 
band projections  reveal that the unoccupied parabolic bands $c1$ and $c2$ in 
Fig.\,\ref{models4b}(b), lying at $\sim$1.1\,eV ($\Gamma$-point) below the 
vacuum level, present no contributions from the atomic orbitals. Instead,  $c1$ 
and $c2$ project on plane-waves characterizing NFE states  spreading out 
(symmetrically) on the oxidized CaO surfaces as shown in  
Fig.\,\ref{models4b}(d). That is,  the fully oxidized \can-BL, CaO/\canii/CaO, 
presents a combination of half-metal channels along  the  \canii\, layers, 
sandwiched by NFE states on top of the oxidized CaO surface.

%%%%%Discussions on the EEF 
% It is worh noting that $c1$ and $c2$ lie at about 2.5\,eV above the Fermi 
% level. Since the NFE states are weakly bonded to the O/\can-BL/O surfaces, the 
% spacial  distribution and energy position of those states can be tuned by 
% external electric field (EEF). In Fig.\,\ref{EEF} we present our results of the 
% energy position and localization of the NFE states as a function of the EEF.
% 
% 
%   \begin{figure}
% %    \includegraphics[width=7.5cm]{fig2/EEF.jpg}
%     \caption{\label{EEF} EEF }
%   \end{figure}

% 
% \begin{table}[h]
% \caption{\label{energy} Energy gain ($\Delta E$) upon spin polarization of 
% \cano.}
% \begin{ruledtabular}
% \begin{tabular}{lcc}
% X    &  x     & xx         \\
% \hline
% x   &   &  \\
% \end{tabular}
% \end{ruledtabular}
% \end{table}
% 

\section{Summary and Conclusions}

We have performed a theoretical study, based on first-principles DFT 
calculations, of the oxidized \can\, electrenes. The structural and atomic 
relaxations of the fully oxidized \can-ML and -BL systems reveal a   
hexagonal\,$\rightarrow$\,tetragonal structural transition, giving rise to CaO 
and CaN (covalently bounded) stacked layers. Those oxidized systems  can viewed 
as two-dimensional  CaO/CaN and CaO/\canii/CaO heterostructures. Both systems 
are metallic, characterized by  NFE states near the vacuum level, and  localized 
above the oxidized CaO surfaces. The latter  exhibits a (mechanically tunable) 
FM phase (strength), where the magnetic moment is mainly localized on the 
nitrogen atoms; giving rise to half-metallic energy bands projected mostly on 
the N-$2p_{x,y}$ orbitals. The structural and thermal stabilities of 
CaC/\canii/CaO have been confirmed  through  phonon spectra  and  molecular 
dynamic calculations. Those findings suggest that  CaO/\canii/CaO, a half-metal 
intercalated by NFE states, is a promising   material to the  development of 
electronic and spintronic nanodevices based on 2D materials.

%\section{Appendix}

\begin{acknowledgments}

The authors acknowledge financial support from the Brazilian agencies CNPq, 
CAPES, and FAPEMIG, and the CENAPAD-SP and Laborat{\'o}rio Nacional de 
Computa{\c{c}}{\~a}o Cient{\'i}fica (LNCC-SCAFMat2) for computer time.

\end{acknowledgments}

%\bibliography{ttf-grap4}

\bibliography{elecXadsorb-3.bib}
\end{document}